# 2D Electrophoresis Gel Image and Diagnosis of a Disease

Gene Kim and MyungHo Kim

Bioinformatics Frontier Inc.
93 B Taylor Ave
East Brunswick, NJ
mkim@biofront.biz

**Abstract**[*]: The process of diagnosing a disease from the 2D gel electrophoresis image is a challenging problem. This is due to technical difficulties of generating reproducible images with a normalized form and the effect of negative stain. In this paper, we will discuss a new concept of interpreting the 2D images and overcoming the aforementioned technical difficulties using mathematical transformation. The method makes use of 2D gel images of proteins in serums and we explain a way of representing the images into vectors in order to apply machine-learning methods, such as the support vector machine.

## §1. Introduction

It is known that, if a virus invades a human body, the immune systems of the body will attempt to suppress or deactivate the virus. As a result, the immune systems produce proteins, commonly known as antibodies, which indicate a possibility of the virus producing protein as well. Therefore, it is a reasonable approach to distinguish patients from normal people by identifying the possible changes of proteins in serums. This fits perfectly well with the principle of the papers, [2] and [3]. One of the most powerful techniques available for classification is the process of numericalization, which represents objects as vectors in a measurable Euclidean Hyperspace.

---

[*] *1991 Mathematical Subject Classification*. Primary I.5, J.3; Secondary I.4.1, I.4.3

Our first assumption is the following:

**Assumption I**

*If there is a disease, then there should be a change in proteins in human body.*

Under the assumption above, we will find a method of representing the proteins of serum. One feasible way is to use the 2D gel method with a proper staining method, such as the standard silver staining.

### §2. 2D gel image and staining methods

2D gel electrophoresis is a method that separates proteins in a 2-dimensional plane by mass and pH of proteins. In order to visualize the distribution of proteins, many researchers use the standard silver stain method, allowing the proteins to be visible in very good resolution, which is effective for identification of proteins even if they are present in small quantities. As is often the case in the most of experiments, there are technical problems we have to remedy and compromise so that the process of numericalization is acceptable and tolerable.

1. When the amount of a certain protein reaches a "threshold", the silver stain density decreases. This phenomenon is known as the negative staining effect.
2. Even if the experiments are performed carefully, there always will be some variations of images. For example, in the image, the same protein will not be in the same position relative to other proteins. In other words, the normalization process of images is needed.

The first problem could be eliminated, by inventing a new stain method in the future[1]. However, since we cannot find a better alternative and the negative effect seems consistent, we cannot help but *accept as it is* and boldly assume that

**Assumption II**

*The negative stain effect in the 2D gel image reflects a status of proteins.*

This assumption will be a foundation of representation as a vector, described in the section 4.

Conceptually we have only to find out a method of representation of each serum, which is reproducible with some tolerable variations. This concept is acceptable in the following sense. When we try to identify some person with his/her photo, even if the picture may not be an exact replica of the person, the job is done successfully most of the time. Whatever qualities we observe, estimate or sample, there are variations in measurements or recording, since everything changes and it is impossible to produce the equivalent results every time. Likewise, there are variations in 2D gel images even under the assumption that all the experiments are perfectly accurate and executed the exactly same way. It could be caused by the status of a donor of serums or experimental setting, however, no matter what the variations would be acceptable. From this, we will state our third assumption:

**Assumption III**

---

[1] If there is no negative silver stain and the densities of stains represent quantity of proteins well, then positions and quantities at the positions can be used as a representation into a vector.

*The proteins of the serum will change, but acceptable enough to observe the difference between normal status and abnormal.*

The 2D gel image (dispersed proteins) of a serum may serve as a "protein print" of its donor. The effect of the second problem could be diminished, by transforming the images with respect to some fixed "reference" proteins. This transformation is an essential part of normalization process, which will be discussed in the section 4.

### §3. Experimental scheme

The scheme designed below is for determining the presence of a disease in a person. However, if we consider a set having two statuses of a disease, since it is about separating one set into two or more, we can apply the same scheme for classifying a disease status.

1. Choose a disease.
2. Obtain serums[2] of the same number of patients and normal people.
3. Obtain images of proteins distributed by 2D gel electrophoresis[3]
4. Normalization: Modify the images by transforming with respect to some fixed reference proteins.
5. Representation: Transform each image into a set of numbers depending on the density of pixels.
6. Each set of numbers represents a vector in a Euclidean space whose dimension is the number of total pixels.

---

[2] The scheme could be applied to tissues as well.

[3] A commercial software such as PDQuest could be useful

7. Apply a machine learning method, such as neural network, SVM (support vector machine), decision tree, and others, for classification of the set of vectors obtained to get a generalized cut-off.[4](See [1] for a brief review of SVM)

### §4. Normalization and Representation

In this section, we will describe the normalization of images and once the normalization is done, two different methods of representation into vectors are explained.

### §4.1. Normalization

Suppose we have two images. To compare and contrast one with the other (i.e. some patterns to distinguish one from another), a careful normalization process is required. The normalization can reduce the inevitable error, the size change of images, caused by routine experiments. To minimize this effect, the method chooses two fixed points as reference points. Then, with respect to the two points, the method expands or reduces the images, by using a mathematical transformation. Finally, the method chooses a rectangular area of the same size from each image.

Let us explain in some details for this normalization.

---

[4]This cut-off is used for diagnosis.

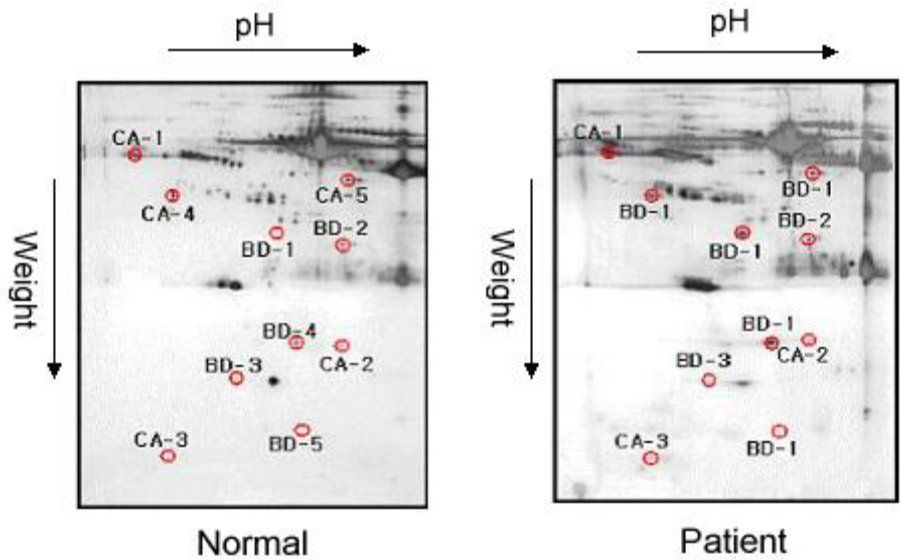

**FIG. 1**

Fig. 1 shows two 2D gel images, one from normal person while the other is from a patient. Although most of proteins change in quantity depending on each person, some of the proteins are always present, as BD-1 and CA-3 appear in both persons.

1. For two acceptable reference points, it is good to consider two spots representing proteins such as BD-1 in Fig. 1 and pick the center point, i.e. a pixel, from the spot of each image.

2. Once the two reference points, say *A* and *B*, are chosen from each image, the method considers coordinate charts on all the images with respect to the number and the position of pixels. Note that the two points are neither on the same horizontal (stretched along pH) nor the same vertical (stretched along weight) line. Thus we have associated coordinates, *x* and *y* to each pixel of each image and a transformation function between image 1 and image 2 may be defined as follows:

$$f: R^2 \to R^2$$

$$f(A_1)=A_2, f(B_1)=B_2$$

where $\{A_1, A_2\}$ and $\{B_1, B_2\}$ are the two reference points in images 1 & 2. The simplest function satisfying these conditions is linear, called an Affine transformation. In mathematical terms, $f(x) = Mx + b$, where $M$ is a two by two matrix and $x$ and $b$ are in $R^2$. The interpolation problem occurs during expansion or reduction, which may be solved by Gauss or linear distribution.

3. Then, the method chooses the area of rectangular form, which is equidistant with respect to the two reference points $A$ and $B$. The number of pixels in each rectangle should be the same for all images.

### §4.2. Representation

Here are two methods for representation of the images for a normalized set of images.

1. Whole rectangle

    Under the assumption II mentioned in the **§2**, take each entire rectangular image and each pixel of it becomes a component of a vector. By enumerating the whole set of numbers corresponding to each pixel in a predetermined order, we will represent an image as a vector in a finite dimensional Euclidean space. Recall that the number of pixels in each image should be fixed.

2. Chosen spots

    Choose a finite number, $K$, of conspicuous spots representing proteins and their quantities, for example, we may take CA-1, BD-1 CA-2 and CA-3. Each of chosen spots has a corresponding number[5], which is the sum of the numbers assigned to each pixel consisting of the spot. Thus, the sum of each spot will

---

[5] Commercial software has a function to compute the density of a spot.

represent the relative quantity of the protein corresponding to the spot relative to other spots. By enumerating the quantities of those four proteins, we have a vector in the four dimensional Euclidean space.

Thus, we obtain a set of vectors representing persons. Suppose we have a group of such vectors and label +1 and −1, depending on whether the person has a disease or not respectively. Then by using SVM(support vector machine), we could find a criterion for diagnosing the disease and the criterion would be used for diagnosis. (For application for clinical data, refer to [1] and for SNP data, see [2] and [3]).

### §5. Discussions

In this section, we discuss two things, choosing a region of an image for representation and perception about the experimental results.

#### §5.1 Comparison of the whole images with chosen spots

At a glance, in representing an electrophoresis image, the second method seems more natural than the first one. However, though considering the quantities of proteins looks like intuitive and appealing to biological meaning, the procedure of measuring the relative quantities of chosen proteins may not be accurate for our purpose, in other words, we might take proteins not related to a disease. On the contrary, accepting the whole image could contain more than we realize. We all know from a meticulous analysis that recognition of a person with a picture is due to the human eye's ability of computing relative positions of specific objects such as nose, eye, mouth, ears, distance between eyes. Each pixel with its own density plays a role as a member of a whole image. Though each pixel does not give any clue by itself, all pixels together with others send us a concrete picture we conceive. Therefore, it is reasonable to say that the intrinsic invariants of an image are the relative position of a pixel with its density. The first method is about considering the whole package of all relative positions and their densities.

### §5.2 Reproducibility of experimental results

There are many researchers, who do not feel comfortable with their experimental results. They said there are too many variations in measurements of multiple gene expression rates in DNAchip, micro-array, 2D gel images etc. nonetheless they made careful procedures in every step. Probably it is due to some preconceptions. Everyday we read and see numbers about heights and weights, but we hardly see anyone who raises a question about the credibility of those numbers, though there must be 1-2 cm, i.e., 10-20 mm variations. If it is measured in meter, the magnitude of variation is notably reduced. Similarly, there is a lot of fluctuation in blood pressure of a single person in a day or even in an hour. From this perception, we might think those measurement methods are delicate and very sensitive, but with belief that the variations are enough to detect a difference between patients and normal. So, what I want to suggest is that we do not need to worry about some variations as long as the experiments are done carefully, but the method of analyzing the data.

### Acknowledement

We are grateful to Profs Chul Ahn of University of Texas at Houston and Larry Shepp at Rutgers University for encouraging comments and criticism. We also give special thanks to Young-Joon Hong, M.D. at Korean Cancer Center Hospital and Prof. Chulwoo Kim at medical school of Seoul National University for invaluable discussions.